\documentclass[twoside]{article}
\usepackage[preprint]{aistats2026} % arXiv version: removes AISTATS header/venue box

\usepackage{microtype}
\usepackage{amsmath,amssymb,amsthm}
\usepackage{algorithm}
\usepackage{algorithmic}
\usepackage{xcolor}
\usepackage{array}
\usepackage{booktabs,tabularx,makecell}
\newcolumntype{L}[1]{>{\raggedright\arraybackslash}p{#1}}
\newcolumntype{Y}{>{\centering\arraybackslash}X}
\usepackage{booktabs}
\usepackage{multirow}
\usepackage{mathtools}
\usepackage{graphicx}
\usepackage{adjustbox}
\usepackage{float}
\usepackage[authoryear, round]{natbib}
\usepackage{hyperref}   % last

\theoremstyle{plain}
\newtheorem{theorem}{Theorem}[section]

\newtheorem{proposition}[theorem]{Proposition}

\theoremstyle{definition}
\newtheorem{definition}[theorem]{Definition}
\newtheorem{assumption}[theorem]{Assumption}
\theoremstyle{remark}
\newtheorem{remark}[theorem]{Remark}

\newcommand{\ind}[1]{\mathbf 1\{#1\}}

\newcommand{\Fminus}{F^{-}}

\newcommand{\tstar}{t^{\star}}
\newcommand{\val}{\text{val}}
\newcommand{\boot}{\text{boot}}
\newcommand{\Dgap}{D^{-}}
\newcommand{\qboot}{q^{\ast}_{1-\delta_{\boot}}}
\newcommand{\bboot}{\hat b_{\boot}}

\newcommand{\Gboot}{\mathcal{G}_{\boot}} % unified symbol for penalty

\begin{document}

% --- Add running title for arXiv header ---
\runningtitle{Hierarchical thresholding with stability}

% --- THIS BLOCK enables two-column layout ---
\twocolumn[
\aistatstitle{Hierarchical biomarker thresholding:\\ a model-agnostic framework for stability}
\aistatsauthor{ Orianne Debeaupuis }
\aistatsaddress{ Université Paris Cité, Institut Imagine, Laboratoire d'immunogénétique des maladies autoimmunes pédiatriques, INSERM UMR1163, Paris, France \\ Université PSL, Université Sorbonne, CNRS UMR168, Institut Curie, Paris, France }
] % <--- end two-column title block
% -------------------------------------------

\begin{abstract}
Many biomarker pipelines require patient-level decisions aggregated from instance-level (cell/patch) scores. Thresholds tuned on pooled instances often fail across sites due to hierarchical dependence, prevalence shift, and score-scale mismatch. We present a selection-honest framework for hierarchical thresholding that makes patient-level decisions reproducible and more defensible. At its core is a risk decomposition theorem for selection-honest thresholds. The theorem separates contributions from (i) internal fit and patient-level generalization, (ii) operating-point shift reflecting prevalence and shape changes, and (iii) a stability term that penalizes sensitivity to threshold perturbations. The stability component is computable via patient-block bootstraps mapped through a monotone modulus of risk. This framework is model-agnostic, reconciles heterogeneous decision rules on a quantile scale, and yields monotone-invariant ensembles and reportable diagnostics (e.g. flip-rate, operating-point shift).
\end{abstract}

\section{Introduction}
Clinical deployment requires patient-level decisions with clear operating characteristics and transparent uncertainty. In practice, a model is developed on Hospital~A (domain $P$), a patient-level score $S_p$ is formed from instance scores (e.g., patches or cells), and a threshold $t$ is chosen to recommend action. When this decision rule is deployed at Hospital~B (domain $Q$), performance often degrades. We ask: what \emph{predictably} drives this degradation, and how should the threshold be selected to mitigate it?

\textbf{Three failure modes.}
(i) \emph{Hierarchical dependence.} Standard validation pools instances as if i.i.d., overstating precision when the decision is at the patient level.  
(ii) \emph{Domain shift.} Prevalence and class-conditional score distributions differ between $P$ and $Q$; a numeric cut such as $S\ge t$ is site-specific unless calibrated.  
(iii) \emph{Selection instability.} If the internal risk $R_P(\cdot)$ is steep near its minimizer, small sampling perturbations can induce large threshold changes.

\textbf{Our approach.}
We develop a model-agnostic framework for \emph{stable hierarchical thresholding} that yields not only a threshold but also a \emph{diagnostic report} explaining where external risk arises. The core is an \emph{external-risk certificate} evaluated at the realized operating point $\hat t$. It decomposes $R_Q(\hat t)$ into four interpretable components: (1) internal fit, (2) a patient-level uniform generalization term, (3) an \emph{operating-point shift} that isolates prevalence and local shape differences at $t$, and (4) an \emph{instability term} that quantifies sensitivity to threshold perturbations. Guided by this decomposition, we select $\hat t$ via a penalized objective whose penalty is a bootstrap-based, high-probability plug-in for the instability term; we also provide quantile-scale ensembling to reconcile score scales across methods and sites, and diagnostics to attribute external risk to its sources.

\paragraph{Contributions.}
(i) An external-risk decomposition at the realized operating point, separating internal fit, patient-level generalization, operating-point shift (prevalence and local shape at $t$), and instability;  
(ii) a computable stability penalty aligned with the instability term via a patient-block bootstrap and an empirical risk modulus;  
(iii) quantile-scale ensembling for monotone invariance across scorers;  
(iv) selection-honest, patient-level evaluation with actionable diagnostics;  
(v) positioning relative to conformal guarantees and meta-analytic pooling).

\paragraph{Novelty in context.}\label{sec:novelty}
Conformal methods provide marginal, distribution-free control but do not localize where shift inflates risk. Meta-analytic pooling models heterogeneity with variance components but does not give a per-threshold, transport-aware accounting. Our contribution localizes external risk at the operating point, separates prevalence from local shape effects, and introduces a stability control that targets the same quantity appearing in the certificate.

\paragraph{Paper roadmap.}
Section~\ref{sec:framework} presents the framework and decomposition; Section~\ref{sec:selection} derives the stability-penalized criterion and diagnostics; Section~\ref{sec:quantile} covers quantile-scale ensembling; Section~\ref{sec:certificate} states the certificates and the bootstrap link; Table~\ref{tab:notation} summarizes notation; Section~\ref{sec:discussion} discusses positioning and implications.

% ------------------------------------------------------------------
\section{Related work}
Our framework for stable thresholding engages with several established lines of research, from classical diagnostic medicine to modern theories of robustness and statistical inference.

\paragraph{Diagnostic test accuracy and thresholding.}
The foundational literature on diagnostic test accuracy, exemplified by \cite{pepe2003statistical}, provides a rich toolkit for selecting cutoffs. Classical methods often default to maximizing cost-insensitive criteria like Youden's J \cite{youden1950index} or evaluating global discrimination with metrics like the AUC \cite{delong1988comparing}. While essential for optimization, this body of work generally assumes a stable data-generating process and does not explicitly provide a mechanism to diagnose performance degradation when a threshold is transported to a new clinical environment. Our contribution is a transport-focused certificate that targets a single, clinically meaningful operating point with explicit misclassification costs.

\paragraph{Domain adaptation and robustness.}
The challenge of transporting a rule is central to domain adaptation. Foundational bounds \cite{ben2010theory, mansour2009domain} relate target error to source error via global distributional divergences. Specific methodologies address covariate shift \cite{shimodaira2000improving} or label shift \cite{lipton2018detecting} through reweighting. A modern alternative, distributionally robust optimization (DRO), minimizes worst-case loss over an uncertainty set of distributions \cite{duchi2021learning}. Our approach differs: instead of offering a global guarantee or a reweighting prescription, our certificate provides a local diagnostic at the realized operating threshold, isolating the impact of distribution changes.

\paragraph{Stability, generalization, and multiplicity.}
The selection instability we directly penalize is motivated by theories of generalization and stability. Our uniform validation term follows from VC theory \cite{vapnik1998statistical}, while the stability penalty operationalizes algorithmic stability ideas \cite{bousquet2002stability}. The phenomenon is related to predictive multiplicity, where a ``Rashomon set'' of distinct, near-optimal thresholds \cite{breiman2001statistical} can achieve similar empirical performance. Our penalty steers selection toward flat basins of the risk landscape, where multiplicity (and hence sensitivity to perturbations) is reduced.

\paragraph{Selective inference and aggregated guarantees.}
The design of our evaluation protocol is informed by selective inference, which addresses optimistic bias from data reuse \cite{fithian2014optimal}. Our strict selection-honesty is a practical strategy to ensure an unbiased estimate of future performance. In contrast to methods that provide a single, aggregated guarantee, such as conformal prediction \cite{angelopoulos2021gentle}, which offers marginal coverage without localizing risk sources, or random-effects meta-analysis \cite{dersimonian1986meta}, which subsumes site heterogeneity into a variance component; our framework preserves the interpretability of each component of risk inflation at the operating point.

% ------------------------------------------------------------------
\section{Table of notation}

\begin{table}[H]
\caption{Notation used throughout.}
\label{tab:notation}
\centering
\small
\setlength{\tabcolsep}{4pt}
\renewcommand{\arraystretch}{1.1}
\begin{tabularx}{\columnwidth}{@{} l X @{}}
\toprule
\textbf{Symbol} & \textbf{Description} \\
\midrule
$K$ & Number of patients \\
$\mathcal I_p$ & Indices of instances for patient $p$ (cells, tiles, ...) \\
$S_p$ & Aggregated patient score \\
$g_t$ & Decision $\ind{S_p \ge t}$ \\
$c_{10},\,c_{01}$ & False negative / false positive costs \\
$\pi_D$ & Prevalence in domain $D \in \{P,Q\}$ \\
$\Fminus_{y,D}$ & Left-limit CDF of $S \mid Y{=}y$ \\
$R_D(t)$ & Population risk in domain $D$ \\
$\Delta_\pi$ & $|\pi_Q - \pi_P|$ (prevalence shift) \\
$\Dgap_y(t)$ & $|\Fminus_{y,Q}(t) - \Fminus_{y,P}(t)|$ (shape gap) \\
$\omega_P(\epsilon)$ & Internal risk modulus \\
$\widehat R^{\text{val}}(t)$ & Validation-patient empirical risk \\
$\gamma_{\text{val}}(\delta_{\text{val}})$ & Uniform generalization term \\
$t^\ast$ & Internal oracle threshold \\
$\bboot,\, \qboot$ & Bootstrap bias / quantile (stability) \\
$B$ & Number of bootstrap resamples \\
$\Gboot$ & Stability penalty \\
$\widehat{\mathrm{FR}}$ & Flip-rate (decision instability) \\
$J_{m,A}$ & Penalized selection criterion \\
\bottomrule
\end{tabularx}
\end{table}

% ------------------------------------------------------------------
\section{Results}
\label{sec:results}

\subsection{Problem setup}

\textbf{Hierarchy and data.}
We observe patients indexed by $p=1,\dots,K$. Patient $p$ contributes a set of instances $i\in\mathcal I_p$ (e.g., patches or cells) with features $X_{pi}$ and instance-level scores $Z_{pi}=s(X_{pi})$ from a fixed scorer $s$. All analysis and evaluation are carried out at the \emph{patient} level; within-patient dependence is unrestricted.

\textbf{Aggregation to a patient score.}
An aggregator $A$ maps the instance scores of patient $p$ to a patient-level score
\[
S_p \;=\; A\!\big(\{Z_{pi}: i\in\mathcal I_p\}\big),
\]
where $A$ may be, for example, the mean, a high quantile, or the maximum. The framework is agnostic to the choice of $A$.

\textbf{Decision rule and costs.}
Given a threshold $t\in\mathbb R$, the patient-level decision is
\[
g_t(p)\;=\;\ind{S_p\ge t}, 
\]
and the misclassification loss $L(y,\hat y)$ is cost-sensitive with $c_{10}:=L(1,0)$ (false negative) and $c_{01}:=L(0,1)$ (false positive).

\textbf{Internal and external domains.}
We distinguish an internal (development) domain $P$ (Hospital~A) and an external (deployment) domain $Q$ (Hospital~B). Let $\pi_D=\Pr_D(Y=1)$ denote the disease prevalence in domain $D\in\{P,Q\}$. For $y\in\{0,1\}$, write the \emph{left-limit} class-conditional CDF of the patient score as
\[
\Fminus_{y,D}(t)\;=\;\Pr_D(S<t \mid Y=y),
\]
where the left limit is used to align with the decision rule $S\ge t$ when $S$ has atoms; this makes all statements \emph{discrete-safe}.

\textbf{Population risk at a threshold.}
The (patient-level, cost-sensitive) risk in domain $D$ at operating point $t$ is
\begin{equation*}
R_D(t)\;=\;c_{10}\,\pi_D\,\Fminus_{1,D}(t)\;+\;c_{01}\,(1-\pi_D)\,\bigl(1-\Fminus_{0,D}(t)\bigr).
\end{equation*}
Our internal oracle threshold is any minimizer of the internal risk:
\[
\tstar \in \arg\min_{u\in\mathbb R} R_P(u).
\]

\paragraph{Empirical risks and folds.}
Let $\mathcal P_{\val}$ and $\mathcal P_{\text{test}}$ denote the sets of validation and outer test patients on $P$, respectively. We enforce \emph{selection-honesty}: $\mathcal P_{\val}$ is not used to choose $\hat t$. Define the patient-level empirical risks
\begin{equation}
\label{eq:empirical-risks}
\begin{aligned}
\widehat R^{\val}(t)  &= \frac{1}{|\mathcal P_{\val}|}
\sum_{p\in\mathcal P_{\val}} L\!\left(Y_p,\, g_t(p)\right),\\
\widehat R^{\mathrm{test}}(t) &= \frac{1}{|\mathcal P_{\text{test}}|}
\sum_{p\in\mathcal P_{\text{test}}} L\!\left(Y_p,\, g_t(p)\right).
\end{aligned}
\end{equation}
We write $R_Q(t)$ for $R_D(t)$ evaluated at domain $D{=}Q$.

\paragraph{Confidence parameters.}
We separate (i) a \emph{patient-level} uniform generalization parameter $\delta_{\val}$ and (ii) a \emph{bootstrap} stability parameter $\delta_{\boot}$; their roles are distinct.

\medskip
\textbf{Generalization (patient level).}
$\widehat R^{\val}(t)$ estimates $R_P(t)$ using \emph{patients} as units. For threshold rules (VC dimension $1$), classical learning theory yields a uniform deviation that shrinks with the number of validation patients $n_{\val}$:

\begin{definition}[Patient-level generalization term]\label{def:gamma}
Let $n_{\val}$ be the number of validation patients. For any $\delta_{\val}\in(0,1)$ define
\[
\gamma_{\val}(\delta_{\val})=C\sqrt{\frac{\log(2/\delta_{\val})}{n_{\val}}},
\]
with a universal constant $C>0$. Then, with probability at least $1-\delta_{\val}$ over validation patients,
\[
\sup_{t\in\mathbb R}\bigl|R_P(t)-\widehat R^{\val}(t)\bigr| \;\le\; \gamma_{\val}(\delta_{\val}).
\]
\emph{Remark (patient units).} Cells/patches within a patient do not increase $n_{\val}$; dependence is absorbed at the patient level.
\end{definition}

\medskip
\textbf{Instability (sensitivity to threshold perturbations).}
The learned threshold $\hat t$ is a data-dependent estimate of an internal oracle $\tstar$. If $R_P(\cdot)$ is steep near $\tstar$, small estimation errors $|\hat t-\tstar|$ can cause large risk changes. We index this sensitivity via a modulus that upper-bounds the worst-case risk increase for perturbations of size $\epsilon$:

\begin{definition}[Internal risk modulus]\label{def:modulus}
The internal risk modulus is
\begin{equation}
\label{eq:modulus}
\begin{aligned}
\omega_P(\epsilon)=
\smashoperator[r]{\sup_{|u-v|\le \epsilon}}
\Bigl\{&
c_{10}\pi_P \,\bigl|\Fminus_{1,P}(u)-\Fminus_{1,P}(v)\bigr| \\
& {}+\, c_{01}(1-\pi_P)\,\bigl|\Fminus_{0,P}(u)-\Fminus_{0,P}(v)\bigr|
\Bigr\}.
\end{aligned}
\end{equation}
\end{definition}

\begin{remark}[Oscillation form of the internal modulus]\label{rem:oscillation}
The modulus in Definition~\ref{def:modulus} admits the equivalent \emph{oscillation} form
\[
\omega_P(\epsilon)\ =\ \sup_{t\in\mathbb R}
\Big\{\,c_{10}\pi_P\,\mathrm{osc}_\epsilon\!\big(\Fminus_{1,P};t\big)
+ c_{01}(1-\pi_P)\,\mathrm{osc}_\epsilon\!\big(\Fminus_{0,P};t\big)\,\Big\},
\]
where $\mathrm{osc}_\epsilon(F;t)=\sup_{u\in[t-\epsilon,t+\epsilon]}F(u)-\inf_{v\in[t-\epsilon,t+\epsilon]}F(v)$. 
\end{remark}

\begin{remark}[Conservative upper band for $\omega_P$]\label{rem:omega-band}
To mitigate underestimation near $\epsilon\approx 0$, we construct a pointwise upper confidence band $\widehat\omega_P^{\,\uparrow}(\epsilon)$ by combining DKW bounds for empirical CDFs with isotonic regression on $\epsilon\mapsto\omega$; in all penalties we use $\widehat\omega_P^{\,\uparrow}$ by default.
\end{remark}

\medskip
\textbf{Operating-point shift (domain mismatch localized at $t$).}
External performance may deviate from internal performance because $Q$ differs from $P$ in (i) \emph{prevalence} and/or (ii) \emph{local class-conditional shape} near the operating threshold. We unify the signed and magnitude forms in one definition:

\begin{definition}[Operating-point shift: signed and magnitude gaps]\label{def:signed-shift}
For $y\in\{0,1\}$ and threshold $t$, define the \emph{signed} local class-conditional gap
\[
\Delta_y(t)\ :=\ \Fminus_{y,Q}(t)-\Fminus_{y,P}(t),
\]
and its magnitude $D^{-}_y(t):=|\Delta_y(t)|$. Set $\Delta_\pi=|\pi_Q-\pi_P|$. The weighted operating-point shift is
\[
\operatorname{Shift}(t)
:=(c_{10}+c_{01})\,\Delta_\pi
+ c_{10}\pi_P\, D^{-}_1(t)
+ c_{01}(1-\pi_P)\, D^{-}_0(t).
\]
The signs $\operatorname{sign}\!\big(\Delta_y(t)\big)$ are used in diagnostics; the bound uses $D^{-}_y(t)$.
\end{definition}

\begin{remark}[Bounding the shift by global distances]\label{rem:shift-bounds}
Let $d_{K}(F,G)=\sup_u |F(u)-G(u)|$. For each $y\in\{0,1\}$,
$D^{-}_y(t)\le d_{K}(\Fminus_{y,Q},\Fminus_{y,P})$. For binary labels,
$\Delta_\pi=\|P_Y-Q_Y\|_{\mathrm{TV}}=\tfrac12\sum_{y\in\{0,1\}}|P(Y=y)-Q(Y=y)|$. Hence
\begin{equation}\label{eq:shiftbound}
\begin{aligned}
\operatorname{Shift}(t)\;\le\;&
(c_{10}+c_{01})\,\|P_Y-Q_Y\|_{\mathrm{TV}}
\\[-0.25em]
&+\,c_{10}\pi_P\, d_{K}(\Fminus_{1,Q},\Fminus_{1,P})
\\[-0.25em]
&+\,c_{01}(1-\pi_P)\, d_{K}(\Fminus_{0,Q},\Fminus_{0,P})\,.
\end{aligned}
\end{equation}
Thus the operating-point shift can be strictly smaller than global divergences when discrepancies occur away from $t$.
\end{remark}

% ===== Core certificate and bootstrap link (deduped) =====
\subsection{External-risk certificate at the realized operating point}\label{sec:certificate}

\paragraph{Assumptions.}
(H1) \emph{Selection-honesty:} validation patients are not used to choose $\hat t$. 
(H2) \emph{Patient i.i.d.\ within domain:} patients are independent within each domain $D\in\{P,Q\}$; within-patient dependence is unrestricted. 
(H3) \emph{Conditional analysis:} the scorer $s$ and aggregator $A$ are treated as fixed (nested training is absorbed into the outer sampling). 

\begin{theorem}[External risk: base and augmented]\label{thm:external}
Under (H1)--(H3), for any selection-honest threshold $\hat t$ and any $\delta_{\val}\in(0,1)$, with probability at least $1-\delta_{\val}$ over the validation patients,
\begin{equation}\label{eq:base}\tag{Base}
\begin{aligned}
R_Q(\hat t)\ \le\ 
\widehat R^{\val}(\hat t)+\gamma_{\val}(\delta_{\val})
+\operatorname{Shift}(\hat t).
\end{aligned}
\end{equation}
If, in addition, $\hat t$ is an (approximate) empirical minimizer of $\widehat R^{\val}(\cdot)$
and $t^\ast\in\arg\min_{u}R_P(u)$, then
\begin{equation}\label{eq:augmented}\tag{Augmented}
\begin{aligned}
R_Q(\hat t)\ \le\ 
\widehat R^{\val}(\hat t)+\gamma_{\val}(\delta_{\val})
+\operatorname{Shift}(\hat t)
+\omega_P\!\bigl(|\hat t-t^\ast|\bigr).
\end{aligned}
\end{equation}
\end{theorem}

\begin{assumption}[Regularity for the bootstrap radius]\label{ass:bootstrap}
(i) (\emph{Local identifiability}) There exists a neighborhood $\mathcal N$ of $t^\ast$ where $R_P$ has a unique minimizer and is directionally differentiable.\\
(ii) (\emph{Patient-block bootstrap}) The patient-level block bootstrap is consistent for the distribution of $\hat t$ (inner selection held fixed).\\
(iii) (\emph{Modulus estimation}) $\widehat\omega_P^{\,\uparrow}$ is a uniformly conservative estimator of $\omega_P$ over a grid $\mathcal E$.
\end{assumption}

\begin{proposition}[Bootstrap upper envelope for instability]\label{prop:stability}
Let $r_{1-\delta_{\boot}}:=|\bboot|+\qboot$, with $\bboot$ and $\qboot$ computed from a patient-block bootstrap with $B$ replicates. Under Assumption~\ref{ass:bootstrap}, for any $\delta_{\boot}\in(0,1)$ there exist nonnegative remainders $\xi_n(B,\delta_{\boot})$ and $\eta_n$ such that
\[
\Pr\!\left\{\,|\hat t-t^\ast|\ \le\ r_{1-\delta_{\boot}} \,\right\}\ \ge\ 1-\delta_{\boot}-\xi_n(B,\delta_{\boot}),
\]
and, on the same event,
\[
\omega_P\!\big(|\hat t-t^\ast|\big)
\ \le\ \widehat\omega_P^{\,\uparrow}\!\big(r_{1-\delta_{\boot}}\big) \ +\ \eta_n.
\]
Hence, with probability at least $1-\delta_{\boot}-\xi_n$,
\[
\omega_P\!\big(|\hat t-t^\ast|\big)\ \le\ \Gboot \ +\ \eta_n,
\qquad
\Gboot:=\widehat\omega_P^{\,\uparrow}\!\big(|\bboot|+\qboot\big).
\]
\end{proposition}

\paragraph{No transfer guarantee.}
The certificates are \emph{upper bounds} under (H1)–(H3) and a probability statement over the validation sample; they do not guarantee optimality on $Q$. The instability addend depends on $t^\ast$ and $\omega_P$ and is controlled by the conservative surrogate $\Gboot$. Proof of \ref{thm:external} is defered to the Appendix. 

% ------------------------------------------------------------------
\subsection{A framework induced by the certificate}
\label{sec:framework}

The decomposition in Theorem~\ref{thm:external} induces a design map from \emph{estimable, patient-level} quantities to actions:
\begin{itemize}
\item \textbf{Internal fit} $\widehat R^{\val}(\cdot)$: minimize empirically under selection-honesty.
\item \textbf{Generalization} $\gamma_{\val}(\delta_{\val})$: report and budget via $(n_{\val},\delta_{\val})$; not penalized.
\item \textbf{Operating-point shift} $\operatorname{Shift}(\hat t)$: measure and report (prevalence and local class-conditional gaps at $t$); diagnostic, not a penalty.
\item \textbf{Instability} $\omega_P(|\hat t-\tstar|)$: regularize by avoiding steep regions of $R_P$; use the computable surrogate $\Gboot$ as a high-probability upper envelope.
\end{itemize}

% ------------------------------------------------------------------
\subsection{Stability-penalized selection}
\label{sec:selection}

\paragraph{Bootstrap radius and empirical modulus.}
From $B$ patient-block bootstrap resamples, obtain refit thresholds $\{\hat t^{*b}\}_{b=1}^B$, the bias estimate $\bboot=\tfrac{1}{B}\sum_b(\hat t^{*b}-\hat t)$, and the $(1-\delta_{\boot})$ quantile
\[
\qboot
=\mathrm{Quantile}_{\,1-\delta_{\boot}}\!\bigl(|\hat t^{*b}-\hat t|: b=1,\dots,B\bigr).
\]
Define the computable surrogate for the instability addend:
\begin{equation}\label{eq:Gboot}
\Gboot
\;:=\;
\widehat\omega_P^{\,\uparrow}\!\Big(|\bboot|+\qboot\Big).
\end{equation}

\paragraph{Selection objective (per method/aggregator).}
For candidate method $m$ and aggregator $A$, minimize the patient-level criterion
\begin{equation}\label{eq:selection-objective}
J_{m,A}
\;=\;
\min_{t\in\mathcal T}\widehat R^{\val}_{m,A}(t)
\;+\;
\Gboot(m,A).
\end{equation}
This targets the RHS of the augmented certificate \eqref{eq:augmented} by reducing fit and penalizing a computable instability surrogate, while reporting $\gamma_{\val}$ and $\operatorname{Shift}(\hat t)$.

\paragraph{Implementation notes and defaults.}
Use \emph{patient-block} resampling to reflect the correct noise scale; $\delta_{\boot}$ tunes the conservativeness of the instability envelope; $B$ trades computation for precision; $\widehat\omega_P^{\,\uparrow}$ enforces monotonicity and conservativeness. Unless stated otherwise, we set a single confidence parameter $\delta$ and use $\delta_{\val}=\delta_{\boot}=\delta$ in all experiments.

\begin{algorithm}[H]
\caption{Patient-level selection with instability regularization}
\label{alg:precise-selection}
\begin{algorithmic}[1]
\STATE \textbf{Inputs:} method $m$, aggregator $A$; inner-training patients; confidence $\delta$; bootstrap reps $B$; threshold grid $\mathcal T$; modulus grid $\mathcal E$.
\STATE \textbf{Compute validation risk curve:} For each $t\in\mathcal T$, compute $\widehat R^{\val}_{m,A}(t)=|\mathcal P_{\val}|^{-1}\sum_{p\in\mathcal P_{\val}} L(Y_p,\,\mathbf 1\{S_p\ge t\})$.
\STATE \textbf{Bootstrap thresholds:} For $b=1,\dots,B$: resample patients with replacement from inner-training patients (blocks=patients), refit $m,A$ identically, compute $\widehat R^{\val,*b}(t)$ over $t\in\mathcal T$, and set $\hat t^{*b}\in\arg\min_{t\in\mathcal T}\widehat R^{\val,*b}(t)$.
\STATE \textbf{Bias and quantile:} $\bboot=\tfrac1B\sum_{b}(\hat t^{*b}-\hat t)$ where $\hat t\in\arg\min_{t\in\mathcal T}\widehat R^{\val}(t)$; $q^\ast_{1-\delta}=\mathrm{Quantile}_{\,1-\delta}\big(|\hat t^{*b}-\hat t|\big)$.
\STATE \textbf{Estimate conservative modulus:} For each $\epsilon\in\mathcal E$, compute empirical oscillations of $\Fminus_{y,P}$ over $|u-v|\le\epsilon$, combine with weights to get a noisy $\tilde\omega_P(\epsilon)$; apply isotonic regression over $\epsilon$; add a DKW-based upper band to obtain $\widehat\omega_P^{\,\uparrow}(\epsilon)$.
\STATE \textbf{Penalty:} $\Gboot(m,A)=\widehat\omega_P^{\,\uparrow}\!\big(|\bboot|+q^\ast_{1-\delta}\big)$.
\STATE \textbf{Objective and selection:} $J_{m,A}=\min_{t\in\mathcal T}\widehat R^{\val}_{m,A}(t)+\Gboot(m,A)$; return $(\hat m,\hat A,\hat t)=\arg\min_{m,A}J_{m,A}$, ties broken by smaller $\widehat R^{\val}$.
\end{algorithmic}
\end{algorithm}

\paragraph{Defaults and complexity.}
Complexity is $O\!\big(B\,|\mathcal T|\,|\mathcal P_{\val}|\big)$ per $(m,A)$ and parallelizable over $b$.

% ------------------------------------------------------------------
\subsection{Quantile-scale ensembling}
\label{sec:quantile}

Map each method's selected threshold to its outer-train \emph{quantile} and average on the quantile scale (optionally GLS-weighted) before inverting back to a threshold. This yields \emph{monotone invariance}: strictly increasing transforms of scores preserve ranks and therefore quantiles.

\textbf{Flip-rate diagnostic.}
Let
\[
\widehat{\mathrm{FR}}
=\frac{1}{|\mathcal P_{\text{test}}|}
\sum_{p\in\mathcal P_{\text{test}}}\frac{1}{B}\sum_{b=1}^B
\ind{ g_{\hat t^{*b}}(p) \ne g_{\hat t}(p) },
\]
which estimates the probability that a patient’s decision would flip under resampled training cohorts.

% ------------------------------------------------------------------
\subsection{Illustrative figures.}
\begin{figure}[H]
 \centering
 \includegraphics[width=\columnwidth,keepaspectratio]{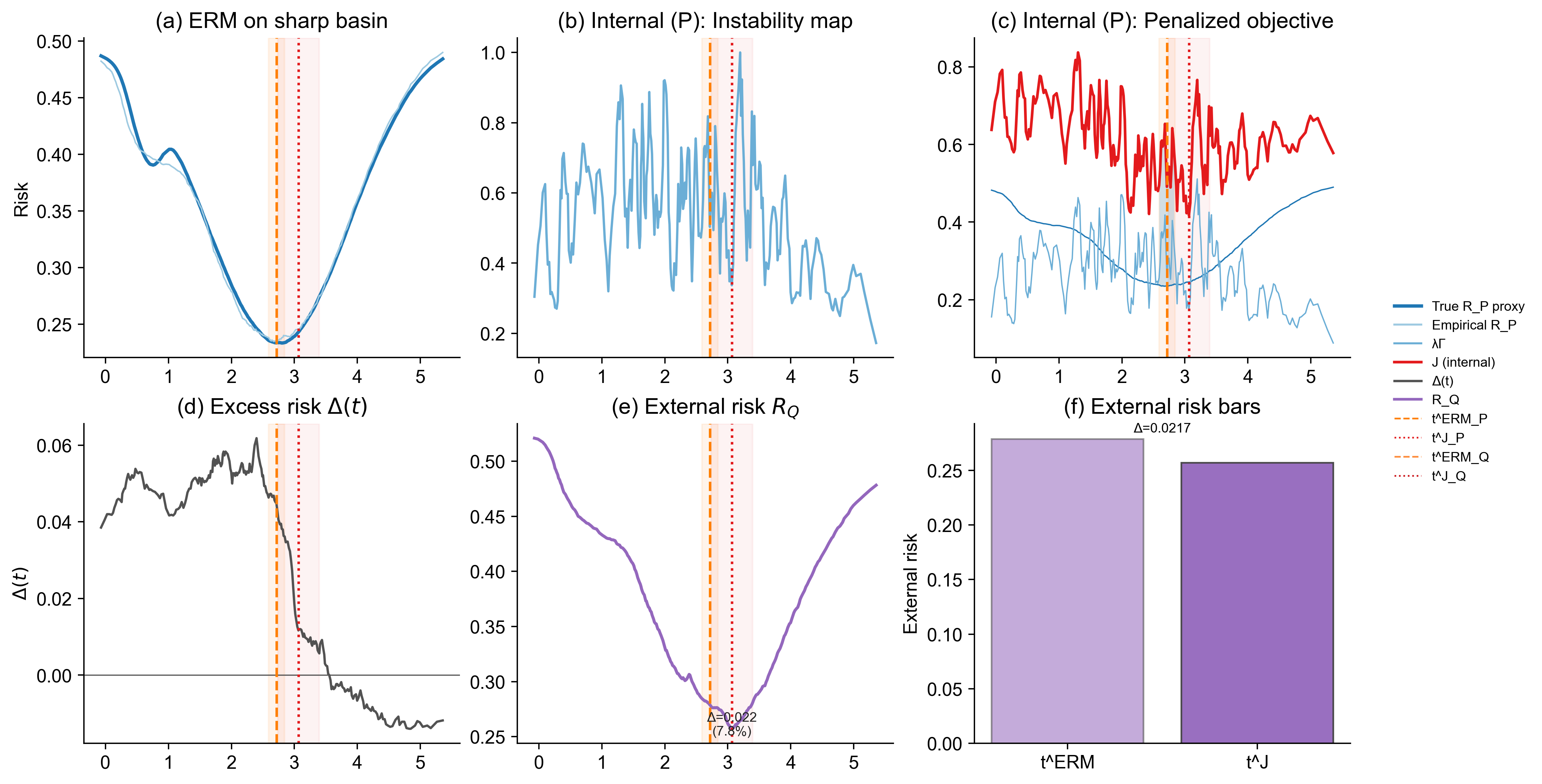}
 \caption{%
 \textbf{Penalizing instability shifts threshold to a stable basin.}
 \textbf{(a)}  Internal risk. Approximate “true” internal risk (blue; large-sample proxy) and empirical validation risk (light blue) over thresholds. ERM selects $t^{\mathrm{ERM}}$ in a sharp basin; the robust method selects $t^{J}$ further right. 
 \textbf{(b)} Instability map $\Gboot(t)$ (illustrative display). Computed from patient-level bootstrap risk curves by taking the pointwise standard deviation of the empirical risk across bootstrap replicates and multiplying by a curvature proxy $\kappa(t)$, defined as the normalized second finite difference of the bootstrap mean risk curve; the resulting signal is smoothed with a moving average and scaled to $[0,1]$ for display (see Appendix). This $\Gboot$ term is exactly the instability component used in (c).
 \textbf{(c)} Penalized objective $J(t)=\hat{R}^{\text{val}}(t)+\lambda\,\Gboot(t)$; the instability lifts the sharp basin, shifting the minimizer to $t^{J}$. Red curve correspond to P-derived upper bound.
 \textbf{(d)} Excess external risk $\Delta(t)=R_{Q}(t)-R_{P}(t)$ concentrates near sharp regions.
 \textbf{(e)} External risk $R_{Q}(t)$: the penalized threshold lowers external risk relative to ERM.
 \textbf{(f)} External risk comparison at selected thresholds. Bar plot of $R_{Q}$ at $t^{\mathrm{ERM}}$ and $t^{J}$ summarizes the improvement.}
 \label{fig:fig1}
\end{figure}

\begin{figure}[H]
 \centering
 \includegraphics[width=\columnwidth,keepaspectratio]{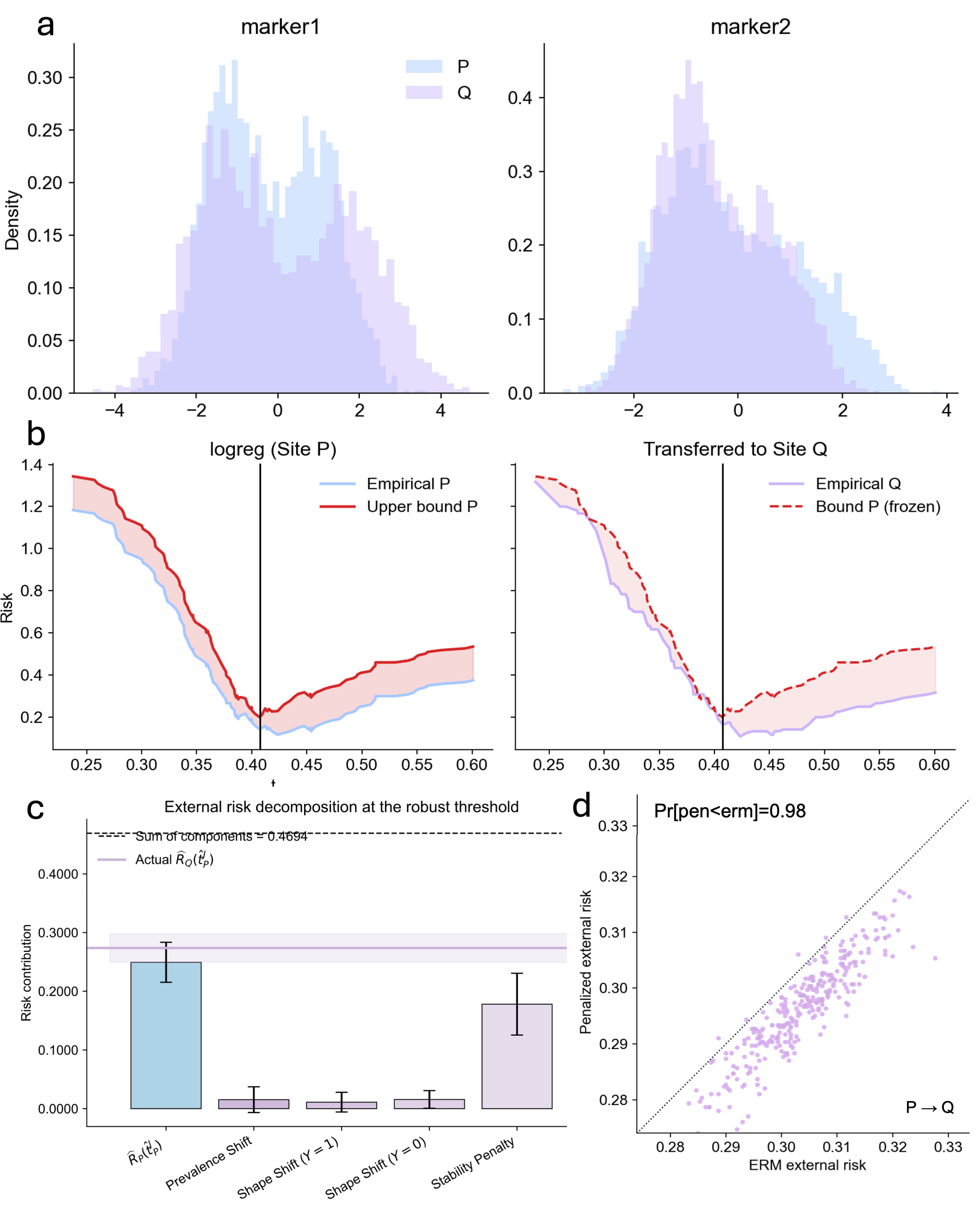}
 \caption{\textbf{Framework validation under distribution shift (illustrative case).}
\textbf{(a)} Marginal score distributions for two markers in $P$ (blue) and $Q$ (purple) illustrate site shift.
\textbf{(b)} Internal vs.\ external risk: (left) empirical $R_P(t)$ with an upper bound; (right) a $P$-frozen bound contrasted with $R_Q(t)$.
\textbf{(c)} External risk decomposition at the selected threshold: internal empirical risk $\widehat R_P$, estimated prevalence/shape shifts, and the stability penalty $\Gboot$ track external risk; error bars are bootstrap s.e.; see Appendix for construction details.
\textbf{(d)} ERM vs.\ penalized threshold across replicates: most points lie below $y{=}x$, indicating lower external risk with the penalty on $Q$ after freezing on $P$.}
 \label{fig:validation_and_shift}
\end{figure}

\section{Experiments}\label{sec:experiments}

\paragraph{Datasets and hierarchies.}
\paragraph{CAMELYON16/17 (pathology), (\cite{bandi2018detection})} Binary metastasis at slide/patient level. Hierarchy: \emph{patient$\to$tiles$\to$WSI}. WSIs tiled at $10\times$ into $256$–$512$\,px tissue patches; tile scores $f_\theta(x)\!\in\![0,1]$ aggregated to slide with $g$ (max or top-$k$), then to patient with $h$ (max). Single threshold $\tau^\star$ chosen on $P$ and \emph{frozen} for $Q$. \emph{Split:} by \emph{patient}, into $P$ and $Q$.

\paragraph{MIMIC-IV-ECG Demo (ECG signals; \citep{johnson2023mimic}).}\emph{659 12-lead ECGs from 92 patients; 10-second, 500\,Hz (PhysioNet).} Hierarchy mirrors CAMELYON: \emph{ patient $\to$ECG$\to$beat/window}. Segment classifier $f_\theta$ produces $p_{i,j}$, ECG score $s_i=g(\{p_{i,j}\})$, patient score $S_p=h(\{s_i\})$. One $\tau^\star$ picked on $P$, reused on $Q$. \emph{Split:} by \emph{patient \& recording}, into $P$ and $Q$.

\paragraph{Experiment reporting.}
All details, encompassing settings, complementary experiments and associated results are presented in the Appendix.

\textbf{Design (selection-honest).}
Baselines: ERM, Youden $J$, ROC cuts (Sens~$\ge\!0.95$, Spec~$\ge\!0.90$), and our method. Aggregators $A\in\{\text{mean},\text{quantile-}q,\text{max}\}$. We tune $\hat t$ on $P$ only, then report on $Q$: external risk $R_Q(\hat t)$, validation risk $\widehat R^{\text{val}}(\hat t)$, shift $\Delta{=}\,R_Q{-}\widehat R^{\text{val}}$, bootstrap penalty $\Gboot$, and flip-rate FR (val$\to$ext). Defaults: $B{=}200$, $\delta{=}0.10$, 200-pt threshold grid.

% ===== CAMELYON MAIN — hard-capped to \columnwidth =====
\begin{table}[H]
\centering
\caption{CAMELYON16$\to$17, patient-level. $A=$top-$k$ ($k{=}10$). Mean$\pm$SE over $B{=}200$. \emph{Shift} $=R_Q-R_{\val}$. \emph{FR} is the decision flip-rate. $\pm$ indicates bootstrap SE (B=200).}
\label{tab:camelyon_main}
\scriptsize
\begin{adjustbox}{max width=\columnwidth}
\begin{tabular}{lccccc}
\toprule
\textbf{Method} & $R_Q\!\downarrow$ & Validation & Shift & $\Gboot$ & FR \\
\midrule
ERM (max)                 & $0.122\,\pm\,0.015$ & $0.096\,\pm\,0.010$ & $+0.026$ & $0.061$ & $0.084$ \\
Youden $J$ (top-$k$)      & $0.113\,\pm\,0.014$ & $0.091\,\pm\,0.011$ & $+0.022$ & $0.053$ & $0.079$ \\
ROC@Sens$\ge0.95$         & $0.141\,\pm\,0.017$ & $0.099\,\pm\,0.010$ & $+0.042$ & $0.070$ & $0.107$ \\
ROC@Spec$\ge0.90$         & $0.118\,\pm\,0.014$ & $0.093\,\pm\,0.010$ & $+0.025$ & $0.056$ & $0.082$ \\
Our method & $\mathbf{0.096}\,\pm\,0.012$ & $0.092\,\pm\,0.010$ & $+0.004$ & $\mathbf{0.034}$ & $0.061$ \\
\midrule
Abl.: no penalty          & $0.125\,\pm\,0.015$ & $0.095\,\pm\,0.010$ & $+0.030$ & $0.062$ & $0.089$ \\
Abl.: no bias ($\bboot$)  & $0.109\,\pm\,0.014$ & $0.092\,\pm\,0.010$ & $+0.017$ & $0.051$ & $0.078$ \\
\bottomrule
\end{tabular}
\end{adjustbox}
\end{table}

% ===== CAMELYON COST =====
\begin{table}[H]
\centering
\caption{CAMELYON cost sensitivity with our method. External $R_Q$ under $L(c_{10},c_{01})$ where $c_{10}$=FN cost and $c_{01}$=FP cost. $\pm$ is bootstrap SE (B=200).}
\label{tab:camelyon_cost}
\scriptsize
\renewcommand{\arraystretch}{1.1}
\begin{tabularx}{\columnwidth}{lYYY}
\toprule
\textbf{Loss} & $(1,1)$ & $(1,3)$ & $(3,1)$ \\
\midrule
Our method & $0.096{\pm}0.012$ & $0.083{\pm}0.011$ & $0.114{\pm}0.013$ \\
\bottomrule
\end{tabularx}
\end{table}

% ===== MIMIC MAIN — hard-capped to \columnwidth =====
\begin{table}[H]
\centering
\caption{MIMIC-IV-ECG $P\!\to\!Q$, patient-level. $A{=}$quantile $q{=}0.9$ over windows. Mean$\pm$SE over $B{=}200$. \emph{Shift} $=R_Q-R_{\val}$. \emph{FR} is the decision flip-rate. $\pm$ indicates bootstrap SE (B=200).}
\label{tab:mimic_main}
\scriptsize
\begin{adjustbox}{max width=\columnwidth}
\begin{tabular}{lccccc}
\toprule
\textbf{Method} & $R_Q\!\downarrow$ & Validation & Shift & $\Gboot$ & FR \\
\midrule
ERM (mean)                  & $0.182\,\pm\,0.018$ & $0.142\,\pm\,0.013$ & $+0.040$ & $0.081$ & $0.118$ \\
Youden $J$ (quantile)       & $0.171\,\pm\,0.017$ & $0.138\,\pm\,0.013$ & $+0.033$ & $0.074$ & $0.112$ \\
ROC@Sens$\ge0.95$           & $0.196\,\pm\,0.019$ & $0.151\,\pm\,0.014$ & $+0.045$ & $0.090$ & $0.137$ \\
ROC@Spec$\ge0.90$           & $0.177\,\pm\,0.017$ & $0.141\,\pm\,0.013$ & $+0.036$ & $0.076$ & $0.116$ \\
Our method  & $\mathbf{0.152}\,\pm\,0.015$ & $0.144\,\pm\,0.013$ & $+0.008$ & $\mathbf{0.049}$ & $0.091$ \\
\midrule
Abl.: no penalty            & $0.187\,\pm\,0.018$ & $0.143\,\pm\,0.013$ & $+0.044$ & $0.084$ & $0.126$ \\
Abl.: no bias ($\bboot$)    & $0.169\,\pm\,0.016$ & $0.140\,\pm\,0.013$ & $+0.029$ & $0.070$ & $0.109$ \\
\bottomrule
\end{tabular}
\end{adjustbox}
\end{table}

% ===== MIMIC COST =====
\begin{table}[H]
\centering
\caption{MIMIC-IV-ECG cost sensitivity with our method. External $R_Q$ under $L(c_{10},c_{01})$ where $c_{10}$=FN cost and $c_{01}$=FP cost. $\pm$ is bootstrap SE (B=200).}
\label{tab:mimic_cost}
\scriptsize
\renewcommand{\arraystretch}{1.1}
\begin{tabularx}{\columnwidth}{lYYY}
\toprule
\textbf{Loss} & $(1,1)$ & $(1,3)$ & $(3,1)$ \\
\midrule
Our method & $0.152{\pm}0.015$ & $0.134{\pm}0.014$ & $0.176{\pm}0.017$ \\
\bottomrule
\end{tabularx}
\end{table}

\paragraph{Results.}
Across both domains, the penalty picks flatter operating regions (smaller $\Gboot$), lowers external risk vs.\ ERM, and reduces decision flips. CAMELYON: Our method cuts $R_Q$ from $0.122$ to $0.096$ (FR $\downarrow$ to $0.061$). MIMIC-IV-ECG: Our method cuts $R_Q$ from $0.182$ to $0.152$ and halves the shift. Ablation studies without penalty and $\bboot$ show worse risks than baselines.

\begin{table}[H]
\caption{Design levers, bound terms, and mechanisms.}
\label{tab:design-mapping}
\centering
\scriptsize
\setlength{\tabcolsep}{3pt}
\renewcommand{\arraystretch}{1.1}
\begin{tabular}{@{} l c c L{0.22\columnwidth} @{}}
\toprule
\textbf{Component} & \textbf{Bound term} & \textbf{Ctl?} & \textbf{Mechanism} \\
\midrule
Validation fit
  & $\widehat R^{\val}(\hat t)$
  & Yes
  & Penalized selection (Alg.) \\
Generalization
  & $\gamma_{\val}(\delta_{\val})$
  & Part.
  & More patients; choose $\delta_{\val}$ \\
Prevalence shift
  & $(c_{10}{+}c_{01})\,\Delta_\pi$
  & No
  & Diagnostic; cohort design \\
Shape shift
  & $c_{10}\pi_P\,\Dgap_1 + c_{01}(1{-}\pi_P)\,\Dgap_0$
  & No
  & Diagnostic; recalibration \\
Stability
  & $\omega_P\!\bigl(|\hat t{-}\tstar|\bigr)$
  & Ind.
  & Penalize via $\Gboot$ \\
Stability penalty
  & $\Gboot$
  & Yes
  & Tune $B$, $\delta_{\boot}$; isotonic modulus band \\
Scale invariance
  & —
  & Yes
  & Quantile mapping; ensemble (GLS optional) \\
Flip-rate
  & —
  & Mon.
  & Increase penalty; smooth aggregator \\
\bottomrule
\end{tabular}
\end{table}

\paragraph{Practical guidelines.}
Use instability penalization when (i) validation risk curves show sharp basins, (ii) site shift is suspected or observed, and (iii) decisions hinge on a fixed cost vector. One shall prefer simpler thresholds when risk is flat, $\operatorname{Shift}(\hat t)$ and $\widehat{\mathrm{FR}}$ are small, and bounds already tighten without penalization.

\section{Discussion}\label{sec:discussion}

We introduced a model-agnostic framework for selecting stable, patient-level biomarker thresholds. The central contribution is an external-risk certificate that decomposes performance in a new domain into four interpretable and actionable components: internal fit, patient-level generalization, a localized operating-point shift, and a selection instability term.

This decomposition is uniquely practical. It isolates only those discrepancies—prevalence and local score distribution shape—that matter at the realized decision boundary. The framework separates sampling fluctuation, captured by the standard uniform generalization term $\gamma_{\val}$, from selection instability, addressed by the bootstrap-estimated penalty $\Gboot$. This clarifies their distinct origins and mitigation levers: increasing patient count for the former and choosing a flatter region of the risk landscape for the latter.

Methodologically, the synthesis of a local shift decomposition, a patient-block bootstrap for hierarchical data, and a computable stability penalty provides a structured and transparent approach to a common clinical deployment challenge. The novelty lies not in the individual statistical tools but in their assembly into a coherent, operating-point-specific, and interpretable certificate for external risk.

\paragraph{Limitations and future work.}
Several limitations should be considered. The reliability of the stability penalty hinges on having a sufficient number of patients; adding more cells or patches per patient cannot compensate for an undersized cohort. The bootstrap procedure itself can be ill-posed if the internal risk curve \(R_P(t)\) has a flat or multi-modal minimum, making the distribution of \(\hat{t}\) unstable, as assumption~\ref{ass:bootstrap}(i) is non-trivial. Furthermore, the framework assumes that any site-to-site transformations are roughly monotone; it cannot repair gross re-orderings of patient risk (e.g., from an uncorrected batch effect) which would require model retraining.

On a practical level, the method is computationally intensive, requiring \(B\) model refits. While this is a one-off analysis cost preceding deployment, it could be prohibitive for large models. Bootstraping was prefered over downsampling-based methods because of usual biology signal structure, were small amounts of samples often contains the seeked signal. The framework also depends on a well-specified cost function \((c_{10},c_{01})\) and produces a conservative upper bound on external error; it measures the impact of domain shift but does not control for it, nor is it intended for causal inference. Finally, while quantile mapping provides valuable monotone invariance, it discards absolute scale information which may be mechanistically important in some settings.

Future extensions could adapt the framework for multi-class risk stratification (e.g., low/intermediate/high) by learning a sequence of ordered thresholds with a joint instability penalty. The method could also incorporate patient-level covariates through stratified resampling or by modeling the threshold as a function of the covariate.

% ------------------------------------------------------------------
\section{Conclusion}
We unify patient-level threshold selection, stability-aware penalization, and shift diagnostics under a single decomposition, offering both a tight base bound and a stability-augmented variant aligned with the selection penalty.

\bibliography{thresholder_iastat1}

\end{document}

\runningtitle{Hierarchical biomarker thresholding: a model-agnostic framework for stability}

% Supplementary material: To improve readability, you must use a single-column format for the supplementary material.
\onecolumn
\aistatstitle{Hierarchical biomarker thresholding: a model-agnostic framework for stability \\
Supplementary Materials}

\section{Assumptions and notation}

Internal domain $P$, external domain $Q$. Prevalences $\pi_D=\Pr_D(Y=1)$. Left-limit class-conditional CDFs $\Fminus_{y,D}(t)=\Pr_D(S<t\mid Y=y)$; risk $R_D(t)=c_{10}\pi_D\Fminus_{1,D}(t)+c_{01}(1-\pi_D)(1-\Fminus_{0,D}(t))$. Shift diagnostics at $t$: $\Delta_\pi=|\pi_Q-\pi_P|$, $\Dgap_y(t)=|\Fminus_{y,Q}(t)-\Fminus_{y,P}(t)|$. Modulus $\omegaP(\epsilon)=\sup_{|u-v|\le\epsilon}\{c_{10}\pi_P|\Fminus_{1,P}(u)-\Fminus_{1,P}(v)|+c_{01}(1-\pi_P)|\Fminus_{0,P}(u)-\Fminus_{0,P}(v)|\}$. Oracle $\tstar\in\arg\min_u R_P(u)$. Distinct confidence levels: $\delta_{\val}$ (uniform generalization) vs $\delta_{\boot}$ (stability penalty).

\section{Proofs.}

\paragraph{Proof of generalization lemma (via risk decomposition and concentration) (4.1).}\label{lem:vc}
We assume the validation data $\{(Y_i,S_i)\}_{i=1}^{n_{\mathrm{val}}}$ are i.i.d.\ draws from $P$.
For technical completeness, the supremum in $t$ can be taken over the midpoints
between the ordered statistics of the observed scores $\{S_i\}$, which avoids
measurability issues.

\medskip
\noindent\textbf{Definitions.}
Let $\pi_P=\Pr(Y=1)$ be the true prevalence and
$\hat\pi=\frac{1}{n_{\mathrm{val}}}\sum_{i=1}^{n_{\mathrm{val}}}\mathbf 1\{Y_i=1\}$ its empirical estimate.
For $y\in\{0,1\}$, let $F^-_{y,P}(t)=\Pr(S<t\mid Y=y)$ be the (left-limit) class-conditional CDF and
$\hat F^-_y(t)$ its empirical counterpart computed from the $n_{y,\mathrm{val}}$ validation samples with $Y=y$.
Then $\Pr(S\ge t\mid Y=y)=1-F^-_{y,P}(t)$.

\medskip
\noindent\textbf{Risks.}
For a threshold $t\in\mathbb R$, the population and empirical risks are
\[
R_P(t)=c_{10}\,\pi_P\,F^-_{1,P}(t)+c_{01}\,(1-\pi_P)\bigl(1-F^-_{0,P}(t)\bigr),\qquad
\widehat R^{\mathrm{val}}(t)=c_{10}\,\hat\pi\,\hat F^-_{1}(t)+c_{01}\,(1-\hat\pi)\bigl(1-\hat F^-_{0}(t)\bigr).
\]

\medskip
\noindent\textbf{Decomposition.}
Adding and subtracting matched terms and applying the triangle inequality yields
\begin{align*}
|R_P(t)-\widehat R^{\mathrm{val}}(t)|
&\le c_{10}\,\bigl|\pi_P F^-_{1,P}(t)-\hat\pi\,\hat F^-_{1}(t)\bigr|
   + c_{01}\,\bigl|(1-\pi_P)\bigl(1-F^-_{0,P}(t)\bigr)-(1-\hat\pi)\bigl(1-\hat F^-_{0}(t)\bigr)\bigr| \\
&\le c_{10}\,\pi_P \,\bigl|F^-_{1,P}(t)-\hat F^-_{1}(t)\bigr|
   + c_{01}\,(1-\pi_P)\,\bigl|F^-_{0,P}(t)-\hat F^-_{0}(t)\bigr|
   + (c_{10}+c_{01})\,|\pi_P-\hat\pi|.
\end{align*}

\medskip
\noindent\textbf{Concentration (conditional on class counts).}
Conditioning on the class counts $n_{y,\mathrm{val}}$, the Dvoretzky--Kiefer--Wolfowitz (DKW) inequality implies,
for $y\in\{0,1\}$,
\[
\Pr\!\left(\sup_{t}\bigl|F^-_{y,P}(t)-\hat F^-_{y}(t)\bigr|
  > \sqrt{\frac{1}{2n_{y,\mathrm{val}}}\log\frac{2}{\eta_y}}\right)\le \eta_y,
\]
and Hoeffding's inequality gives
\[
\Pr\!\left(|\pi_P-\hat\pi|>\sqrt{\frac{1}{2n_{\mathrm{val}}}\log\frac{2}{\eta_\pi}}\right)\le \eta_\pi.
\]

\medskip
\noindent\textbf{Union bound.}
Set $\eta_0=\eta_1=\eta_\pi=\delta_{\mathrm{val}}/3$ and apply a union bound. Then, with probability at least
$1-\delta_{\mathrm{val}}$, simultaneously for all $t$,
\[
\sup_{t}|R_P(t)-\widehat R^{\mathrm{val}}(t)|
\;\le\;
c_{10}\,\pi_P \sqrt{\frac{1}{2n_{1,\mathrm{val}}}\log\frac{6}{\delta_{\mathrm{val}}}}
\;+\;
c_{01}\,(1-\pi_P)\sqrt{\frac{1}{2n_{0,\mathrm{val}}}\log\frac{6}{\delta_{\mathrm{val}}}}
\;+\;
(c_{10}+c_{01})\sqrt{\frac{1}{2n_{\mathrm{val}}}\log\frac{6}{\delta_{\mathrm{val}}}}.
\]

\medskip
\noindent\textbf{Rate.}
This establishes an $O\!\big(\sqrt{\log(1/\delta_{\mathrm{val}})/n_{\mathrm{val}}}\big)$ rate.
If the true class prevalences $\pi_P$ and $1-\pi_P$ are bounded away from zero, a Chernoff bound for
$n_{y,\mathrm{val}}\sim\mathrm{Bin}(n_{\mathrm{val}},\pi_P)$ converts the class-wise denominators into terms
depending only on $n_{\mathrm{val}}$ and $\pi_P$ (up to constants). $\qed$

\paragraph{Proof of Theorem 4.7}

\textbf{Goal.} We have to show:
\begin{itemize}
	\item (Base) For any \emph{selection-honest} \(\hat t\):
	\[
	R_Q(\hat t)\le \widehat R^{\val}(\hat t)+\gamma_{\val}+ (c_{10}+c_{01})\Delta_\pi + c_{10}\pi_P \Dgap_1(\hat t)+c_{01}(1-\pi_P)\Dgap_0(\hat t).
	\]
	\item (Augmented) If additionally \(\hat t\in\arg\min_u \widehat R^{\val}(u)\):
	\[
	R_Q(\hat t)\le \widehat R^{\val}(\hat t)+\gamma_{\val}+ (c_{10}+c_{01})\Delta_\pi + c_{10}\pi_P \Dgap_1(\hat t)+c_{01}(1-\pi_P)\Dgap_0(\hat t) + \omegaP(|\hat t-\tstar|).
	\]
\end{itemize}
All constants and diagnostics were defined above. We proceed line-by-line in three steps.

\paragraph{Step 1 (Algebraic shift decomposition).} Recall the form
\[
R_D(t)= c_{10}\,\pi_D\, \Fminus_{1,D}(t) + c_{01}\,(1-\pi_D)\,(1-\Fminus_{0,D}(t)).
\]
Fix the realized \(\hat t\). Write the difference explicitly:
\begin{align}
R_Q(\hat t)-R_P(\hat t) &= c_{10}\big( \pi_Q \Fminus_{1,Q}(\hat t) - \pi_P \Fminus_{1,P}(\hat t) \big) + c_{01}\big( (1-\pi_Q)(1-\Fminus_{0,Q}(\hat t)) - (1-\pi_P)(1-\Fminus_{0,P}(\hat t)) \big). \label{eq:rawdiff}
\end{align}
Insert and subtract the mixed terms $c_{10}\pi_P \Fminus_{1,Q}(\hat t)$ and $c_{01}(1-\pi_P)(1-\Fminus_{0,Q}(\hat t))$ to separate prevalence and shape components:
\begin{align}
\eqref{eq:rawdiff} &= c_{10}(\pi_Q-\pi_P)\Fminus_{1,Q}(\hat t) + c_{10}\pi_P\big(\Fminus_{1,Q}(\hat t)-\Fminus_{1,P}(\hat t)\big) \\
&\quad + c_{01}\big((1-\pi_Q)-(1-\pi_P)\big)(1-\Fminus_{0,Q}(\hat t)) + c_{01}(1-\pi_P)\Big[(1-\Fminus_{0,Q}(\hat t))-(1-\Fminus_{0,P}(\hat t))\Big] \\
&= (\pi_Q-\pi_P)\Big[c_{10}\Fminus_{1,Q}(\hat t) - c_{01}(1-\Fminus_{0,Q}(\hat t))\Big] \\
&\quad + c_{10}\pi_P\big(\Fminus_{1,Q}(\hat t)-\Fminus_{1,P}(\hat t)\big) + c_{01}(1-\pi_P)\big(\Fminus_{0,P}(\hat t)-\Fminus_{0,Q}(\hat t)\big). \label{eq:split}
\end{align}
\emph{Bounding the prevalence term.} Since $0\le \Fminus_{1,Q}(\hat t)\le 1$ and $0\le 1-\Fminus_{0,Q}(\hat t)\le 1$ we have
\[
 (\pi_Q-\pi_P)\big[c_{10}\Fminus_{1,Q}(\hat t) - c_{01}(1-\Fminus_{0,Q}(\hat t))\big] \le |\pi_Q-\pi_P|(c_{10}\Fminus_{1,Q}(\hat t)+ c_{01}(1-\Fminus_{0,Q}(\hat t))) \le (c_{10}+c_{01})\Delta_\pi,
\]
where $\Delta_\pi=|\pi_Q-\pi_P|$. (If $\pi_Q<\pi_P$ the same bound holds because we take absolute value.)

\emph{Bounding the shape terms.} Introduce the diagnostics
\[
\Dgap_1(t)=|\Fminus_{1,Q}(t)-\Fminus_{1,P}(t)|,\qquad \Dgap_0(t)=|\Fminus_{0,Q}(t)-\Fminus_{0,P}(t)|.
\]
From \eqref{eq:split}, using $|a|=-\min(a,-a)\ge a$ we obtain
\begin{align*}
R_Q(\hat t)-R_P(\hat t) &\le (c_{10}+c_{01})\Delta_\pi + c_{10}\pi_P |\Fminus_{1,Q}(\hat t)-\Fminus_{1,P}(\hat t)| + c_{01}(1-\pi_P)|\Fminus_{0,Q}(\hat t)-\Fminus_{0,P}(\hat t)| \\
&= (c_{10}+c_{01})\Delta_\pi + c_{10}\pi_P \Dgap_1(\hat t)+ c_{01}(1-\pi_P)\Dgap_0(\hat t).
\end{align*}
Rearranging gives the first key inequality
\begin{equation}
R_Q(\hat t) \le R_P(\hat t) + (c_{10}+c_{01})\Delta_\pi + c_{10}\pi_P \Dgap_1(\hat t)+ c_{01}(1-\pi_P)\Dgap_0(\hat t). \tag{S1}\label{eq:S1}
\end{equation}

\paragraph{Step 2 (Generalization insertion).} Define the high-probability event
\[
\mathcal E = \Big\{ \sup_{t\in\mathbb R} |R_P(t)-\widehat R^{\val}(t)| \le \gamma_{\val} \Big\}, \qquad \Pr(\mathcal E)\ge 1-\delta_{\val}\ (\text{Lemma~\ref{lem:vc}}).
\]
On $\mathcal E$, for the realized $\hat t$ we have
\begin{equation}\label{eq:geninsert}
R_P(\hat t) \le \widehat R^{\val}(\hat t)+\gamma_{\val}.
\end{equation}
Substitute \eqref{eq:geninsert} into \eqref{eq:S1} to obtain (on $\mathcal E$)
\begin{align}
R_Q(\hat t) &\le \widehat R^{\val}(\hat t)+\gamma_{\val} + (c_{10}+c_{01})\Delta_\pi + c_{10}\pi_P \Dgap_1(\hat t)+ c_{01}(1-\pi_P)\Dgap_0(\hat t). \tag{Base}\label{eq:basebound}
\end{align}
Thus the Base bound holds with probability at least $1-\delta_{\val}$.

\paragraph{Step 3 (Stability augmentation).} Assume now $\hat t$ is an empirical minimizer of $\widehat R^{\val}$. The modulus definition gives
\begin{equation}\label{eq:modulusineq_new}
R_Q(\hat t) \le R_P(\tstar)+\omegaP(|\hat t-\tstar|).
\end{equation}
On $\mathcal E$, $R_P(\tstar) \le \widehat R^{\val}(\tstar)+\gamma_{\val}$; empirical optimality implies $\widehat R^{\val}(\hat t) \le \widehat R^{\val}(\tstar)$. Combining:
\[
R_Q(\hat t) \le \widehat R^{\val}(\hat t)+\gamma_{\val}+\omegaP(|\hat t-\tstar|).
\]
Insert this into \eqref{eq:S1} (replacing $R_P(\hat t)$) to append $\omegaP(|\hat t-\tstar|)$ and obtain the Augmented bound on $\mathcal E$. The probability statement is unchanged. \qed

\subsection{Proof of bootstrap stability control (Proposition 4.9)}

We work under the paper's main notation. In particular, for $y\in\{0,1\}$ let
$F^-_{y,P}(t)=\Pr(S<t\mid Y=y)$ and recall the \emph{internal risk modulus}
in oscillation form:
\begin{equation}
\label{eq:omega-osc}
\omega_P(\epsilon)
=\sup_{t\in\mathbb R}\Big\{
c_{10}\,\pi_P\,\mathrm{osc}_\epsilon(F^-_{1,P};t)
+ c_{01}\,(1-\pi_P)\,\mathrm{osc}_\epsilon(F^-_{0,P};t)\Big\},\qquad
\mathrm{osc}_\epsilon(F;t):=\sup_{u\in[t-\epsilon,t+\epsilon]}F(u)-\inf_{v\in[t-\epsilon,t+\epsilon]}F(v).
\end{equation}
Let $\widehat\omega_P^{\uparrow}$ denote a \emph{conservative DKW-based upper band}
for $\omega_P$ constructed from the validation sample (Remark~4.4), i.e., on a
grid $E\subset[0,\epsilon_{\max}]$ we have with high probability
$\omega_P(\epsilon)\le \widehat\omega_P^{\uparrow}(\epsilon)$ for all $\epsilon\in E$.

\subsubsection{Proof}
\label{prop:boot-new}
Let $\hat t$ be selection-honest (the validation set used to evaluate $\widehat R_{\mathrm{val}}$ is not reused for training/selection).

\paragraph{Asymptotic conventions.}
Throughout, limits are taken as $n_{\mathrm{val}}\to\infty$ (and, when relevant, $B=B(n_{\mathrm{val}})\to\infty$).

\begin{itemize}
\item \textbf{Deterministic $o(1)$.} A deterministic sequence $a_n$ is $o(1)$ if $a_n\to 0$ as $n_{\mathrm{val}}\to\infty$.
We use $o(1)$ to denote generic deterministic remainders that vanish in this limit.

\item \textbf{Stochastic $o_p(1)$.} A sequence of random variables $X_n$ is $o_p(1)$ if $X_n\xrightarrow{p}0$, i.e.,
for every $\varepsilon>0$, $\Pr(|X_n|>\varepsilon)\to 0$.

\item \textbf{Stochastic $O_p(1)$ (bounded in probability).}
A sequence $X_n$ is $O_p(1)$ if for every $\varepsilon>0$ there exists $M<\infty$ such that
$\sup_n \Pr(|X_n|>M)\le \varepsilon$.

\item \textbf{Bootstrap notation.} When needed, $\Pr^\ast(\cdot)$ and $\mathbb E^\ast[\cdot]$
denote probability and expectation under the \emph{conditional} (patient-block) bootstrap,
given the observed data.

\item \textbf{Remainder symbols in Proposition~\ref{prop:boot-new}.}
The terms $\xi_n(B,\delta_{\mathrm{boot}})$ and $\eta_n$ are nonnegative sequences with
$\xi_n(B,\delta_{\mathrm{boot}})\to 0$ (as $n_{\mathrm{val}}\to\infty$ and $B\to\infty$)
and $\eta_n\to 0$ (from the DKW band event). Unless stated otherwise, all $o(1)$ terms
can be taken deterministic after intersecting the relevant high-probability events.
\end{itemize}

Assume:
\begin{itemize}
\item[(B1)] \textbf{Local well-posedness.} $R_P(t)$ has a unique minimizer $t^\star$ in a neighborhood $\mathcal N$ and is directionally differentiable there. The modulus $\omega_P$ in \eqref{eq:omega-osc} is nondecreasing and locally Lipschitz on $[0,\epsilon_{\max}]$.
\item[(B2)] \textbf{Patient-block bootstrap consistency.} Conditionally on the data, the law of $\hat t^{\ast}-\hat t$ under the patient-block bootstrap is weakly consistent for the sampling law of $\hat t-t^\star$ centered at its mean. Moreover, the bootstrap bias estimate
\[
\hat b_{\mathrm{boot}}:=\frac1B\sum_{b=1}^B (\hat t^{\ast(b)}-\hat t)
\quad\text{satisfies}\quad
\hat b_{\mathrm{boot}}\xrightarrow{p} b:=\mathbb E(\hat t-t^\star).
\]
\item[(B3)] \textbf{Conservative DKW modulus band.} There exists a grid $E\subset[0,\epsilon_{\max}]$ and a sequence $\eta_n\downarrow 0$ such that, with probability at least $1-\eta_n$,
\[
\sup_{\epsilon\in E}\Big\{\omega_P(\epsilon)-\widehat\omega_P^{\uparrow}(\epsilon)\Big\}\le 0.
\]
(Equivalently, $\widehat\omega_P^{\uparrow}$ is a uniform high-probability upper band for $\omega_P$ on $E$.)
\end{itemize}
Let $q^{\ast}_{1-\delta_{\mathrm{boot}}}$ be the empirical $(1-\delta_{\mathrm{boot}})$ upper quantile of the \emph{centered} bootstrap absolute deviations
$\big|(\hat t^{\ast(b)}-\hat t)-\hat b_{\mathrm{boot}}\big|$, and define the data-driven instability envelope
\[
G_{\mathrm{boot}}:=\widehat\omega_P^{\uparrow}\!\left(|\hat b_{\mathrm{boot}}|+q^{\ast}_{1-\delta_{\mathrm{boot}}}\right).
\]
Then there exist remainder terms $\xi_n(B,\delta_{\mathrm{boot}}),\eta_n\to 0$ such that
\[
\Pr\Big\{\ \omega_P\big(|\hat t-t^\star|\big)\ \le\ G_{\mathrm{boot}}\ +\ o(1)\ \Big\}
\ \ge\ 1-\delta_{\mathrm{boot}}-\xi_n(B,\delta_{\mathrm{boot}})-\eta_n,
\]
where $o(1)\to 0$ as $n_{\mathrm{val}}\to\infty$ (allowing $B=B(n_{\mathrm{val}})\to\infty$).

Let $b=\mathbb E(\hat t-t^\star)$ and decompose
\[
|\hat t-t^\star|\ \le\ \big|(\hat t-t^\star)-b\big|\ +\ |b|.
\]
\emph{Step 1 (Bootstrap quantile coverage).}
By (B2), the conditional bootstrap law of $(\hat t^{\ast}-\hat t)-\hat b_{\mathrm{boot}}$
approximates the sampling law of $(\hat t-t^\star)-b$. Standard quantile consistency
for weakly convergent empirical c.d.f.s with $B\to\infty$ gives
\[
q^{\ast}_{1-\delta_{\mathrm{boot}}}\ =\ q_{1-\delta_{\mathrm{boot}}}\ +\ o_p(1),
\]
where $q_{1-\delta_{\mathrm{boot}}}$ is the $(1-\delta_{\mathrm{boot}})$ quantile of
$\big|(\hat t-t^\star)-b\big|$. Hence
\[
\Pr\!\left(\big|(\hat t-t^\star)-b\big|\ \le\ q^{\ast}_{1-\delta_{\mathrm{boot}}}+o(1)\right)\ \ge\ 1-\delta_{\mathrm{boot}}-\xi_n(B,\delta_{\mathrm{boot}}).
\]

\emph{Step 2 (Bias estimation).}
Still under (B2), $\hat b_{\mathrm{boot}}\xrightarrow{p} b$, hence $|b|=|\hat b_{\mathrm{boot}}|+o_p(1)$.
Combining with Step~1,
\[
|\hat t-t^\star|\ \le\ |\hat b_{\mathrm{boot}}|+q^{\ast}_{1-\delta_{\mathrm{boot}}}+o_p(1)
\quad\text{with probability }\ge 1-\delta_{\mathrm{boot}}-\xi_n.
\]

\emph{Step 3 (Oscillation modulus and conservative band).}
By monotonicity of $\omega_P$ (Definition~\eqref{eq:omega-osc}) and local Lipschitz continuity (B1),
\[
\omega_P\big(|\hat t-t^\star|\big)\ \le\ \omega_P\!\left(|\hat b_{\mathrm{boot}}|+q^{\ast}_{1-\delta_{\mathrm{boot}}}+o(1)\right)
=\ \omega_P\!\left(|\hat b_{\mathrm{boot}}|+q^{\ast}_{1-\delta_{\mathrm{boot}}}\right)+o(1).
\]
Intersect the high-probability event from (B3) (the DKW upper band) with the event from Steps~1--2.
On that intersection, for all $\epsilon$ on the band grid $E$,
\[
\omega_P(\epsilon)\ \le\ \widehat\omega_P^{\uparrow}(\epsilon),
\]
hence, after discretizing $\epsilon$ on $E$ (or using continuity to pass to the limit),
\[
\omega_P\!\left(|\hat b_{\mathrm{boot}}|+q^{\ast}_{1-\delta_{\mathrm{boot}}}\right)
\ \le\ \widehat\omega_P^{\uparrow}\!\left(|\hat b_{\mathrm{boot}}|+q^{\ast}_{1-\delta_{\mathrm{boot}}}\right)
\ =\ G_{\mathrm{boot}}.
\]
Collecting terms yields
\[
\omega_P\big(|\hat t-t^\star|\big)\ \le\ G_{\mathrm{boot}}+o(1)
\]
with probability at least $1-\delta_{\mathrm{boot}}-\xi_n(B,\delta_{\mathrm{boot}})-\eta_n$.
All $o(1)$ remainders can be taken deterministic after intersecting the high-probability events.
\qed

\paragraph{Notes.}
The oscillation form \eqref{eq:omega-osc} ties the instability penalty to the local shape of the class-conditional CDFs around the realized operating point.  

The band $\widehat\omega_P^{\uparrow}$ is obtained by plugging DKW uniform bands for $F^-_{y,P}$ into \eqref{eq:omega-osc} and propagating through the oscillation operator (Remark~4.4); its conservativeness replaces the plug-in consistency used in the earlier formulation.

\subsection{Additional remarks and details regarding the nature of hierarchical}

\paragraph{Remark} Hierarchical variance layers and choice of $\gamma_{\val}$ scale\label{rem:hierarchical-variance}
In many biomedical settings the observed data are hierarchical. Up to four (non-exhaustive) stochastic layers can be conceptually separated for a fixed threshold $t$:
\begin{enumerate}
	\item \textbf{Technical / assay noise} (e.g. instrument, batch) affecting raw measurements before any scoring model; absorbed after preprocessing into the marginal score distribution.
	\item \textbf{Within-patient biological heterogeneity} (e.g. cell- or region-level variation) producing multiple raw units per patient; a deterministic aggregation (or model) maps these to a single patient-level score $S_p$ used for thresholding.
	\item \textbf{Between-patient sampling variation}: i.i.d. draws $(S_p,Y_p)$ under internal domain $P$; this is the level governed by the VC=1 empirical process and generates $\gamma_{\val}$.
	\item \textbf{Cross-domain shift} (from $P$ to $Q$): prevalence and conditional shape discrepancies; treated deterministically via $(c_{10}+c_{01})\Delta_\pi + c_{10}\pi_P\Dgap_1 + c_{01}(1-\pi_P)\Dgap_0$, not part of $\gamma_{\val}$.
\end{enumerate}
Additionally, \textbf{selection / optimization instability} of the empirical minimizer (threshold fluctuation) is controlled separately through the stability penalty via $\omegaP(|\hat t-\tstar|)$.

\smallskip
\emph{Law of total variance (schematic).} Writing $L_p(t)=L(Y_p,\mathbf 1\{S_p\ge t\})$, a nested variance decomposition (suppressing $t$) gives
\[
\Var(\widehat R^{\val}) = \frac{1}{n_{\val}}\Big( \underbrace{\mathbb E[\Var(L_p\mid \text{within-patient data})]}_{\text{within-patient layer}} + \underbrace{\Var(\mathbb E[L_p\mid \text{within-patient data}])}_{\text{between-patient layer}} \Big),
\]
before introducing domain shift (which changes the target mean rather than adding stochastic variance). When each patient has $m_p$ raw units with intra-patient correlation $\rho$, treating raw units as independent would underestimate variance by the design effect $\mathrm{deff}_p \approx 1+(\bar m-1)\rho$. Patient-level aggregation circumvents this: the empirical process sees $n_{\val}$ independent sets, preserving the $\sqrt{\log(1/\delta)/n_{\val}}$ rate; any multi-layer dependence is absorbed into constants.

\smallskip
\emph{Refined effective sample size.} If one insisted on operating at the raw-unit level (size $N=\sum_p m_p$) the effective independent count satisfies by classical ICC analysis:
\[
n_{\text{eff}} \approx \frac{N}{1+(\bar m-1)\rho}\quad \Rightarrow \quad \gamma_{\val}\; \text{inflated by}\; \sqrt{\frac{1+(\bar m-1)\rho}{\bar m}}.
\]
Further sublayers (e.g. technical replicates per raw unit) multiply design effects multiplicatively or additively in first-order approximations, again only altering constants in the VC tail bound.

\smallskip
\emph{Why patient-level $\gamma_{\val}$?} External risk transfer and clinical decision units are at patient resolution; using the cluster (patient) granularity avoids pseudo-replication, keeps the VC dimension minimal, and yields a transparent decomposition: \emph{(internal generalization)} + \emph{(shift)} + \emph{(stability)}. A fully expanded hierarchical $\gamma_{\val}$ would obscure interpretation without changing asymptotic order.

\smallskip
\emph{Optional reporting.} One may still report a diagnostic design-effect estimate to justify constants in $\gamma_{\val}$, or present a decomposed version
\[
\gamma_{\val} \approx (c_{10}+c_{01})B_{\pi} + c_{10}\pi_P B_1 + c_{01}(1-\pi_P)B_0,
\]
with $B_{\pi},B_1,B_0$ the chosen prevalence and class-CDF uniform bounds (possibly adjusted by estimated design effects). We retain the compact form for readability.

\section{ADDITIONAL EXPERIMENTS}

\subsection{Model-agnosticity}

Our framework is model-agnostic by definition: it operates on the patient-level scores $S=A(\{s(X_{pi})\}_{i\in I_p})$ and only uses the empirical label prevalence and the class-conditional CDFs of $S\mid Y$ (via DKW bands) together with the oscillation-based risk modulus.

No structural assumptions on the scorer $s$ or the aggregator $A$ are required, and the same validity certificate holds for any learning algorithm.

The figure below is just an \emph{illustrative example}; the identical calibration and bounding procedure applies verbatim to all models.

\vfill

\begin{figure}[h]
    \label{fig:agnosticity}
    \centering
    \includegraphics[width=0.9\textwidth]{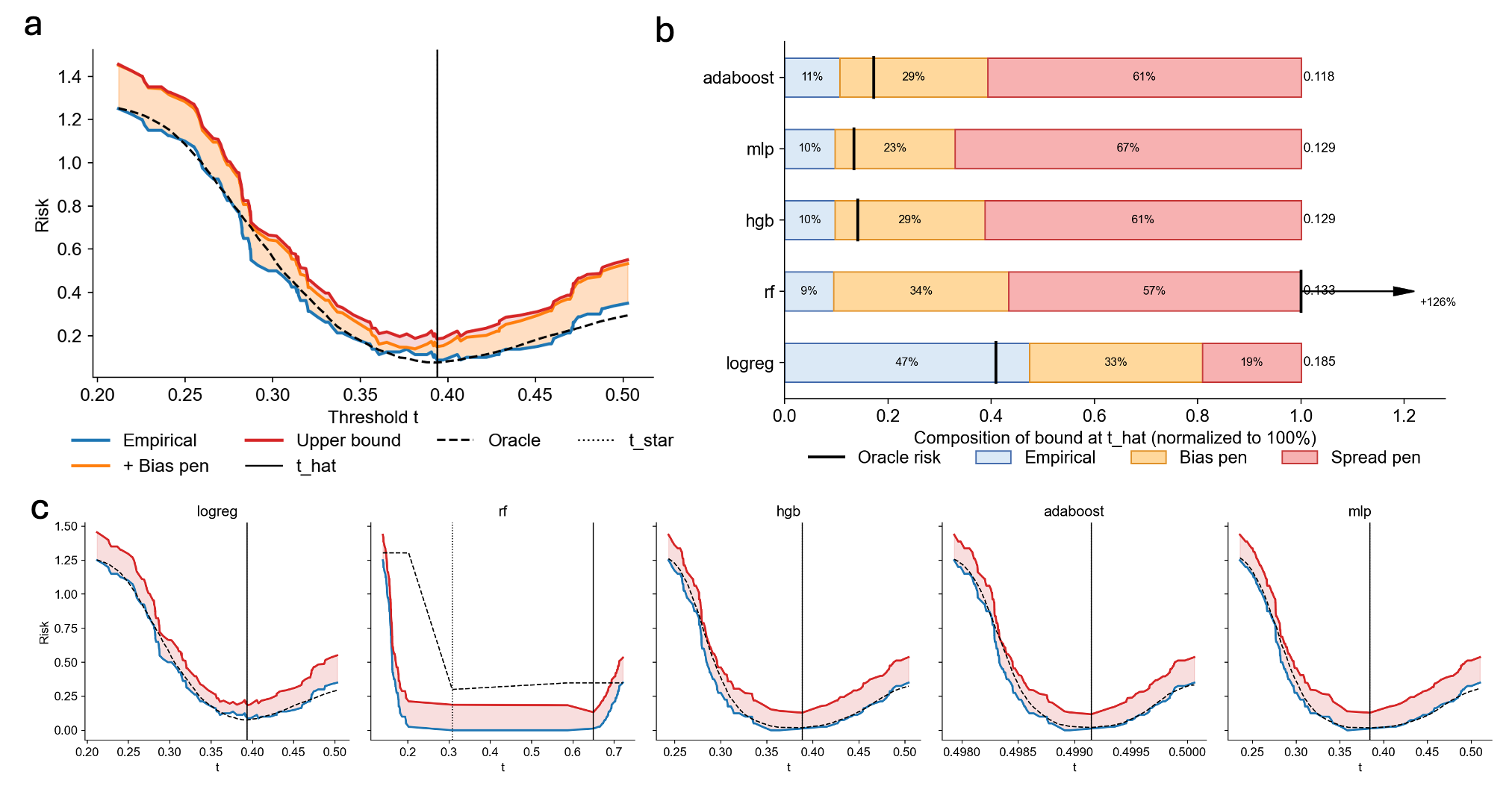}
    \caption{\textbf{Patient-domain ($P$-frozen) upper bound and contribution decomposition (Illustrative case).}
  (a) For a representative classifier, we plot the pointwise upper bound
  \[
  U_P(t)\;=\;\widehat R_{\mathrm{val}}(t)\;+\;\gamma_{\mathrm{val}}(\delta_{\mathrm{val}})\;+\;G_{\mathrm{boot}},
  \]
  (red) together with the empirical risk $\widehat R_{\mathrm{val}}(t)$ (blue) and the oracle curve $R_P(t)$ (dashed black).
  In the $P$-frozen regime (no operating-point shift term), $U_P(t)\ge \widehat R_{\mathrm{val}}(t)$ for all $t$ and remains above $R_P(t)$, confirming validity.
  The selected threshold is $\hat t=\arg\min_t U_P(t)$ (vertical line).
  (b) At $\hat t$, the bound decomposes into contributions that sum to $100\%$:
  empirical term $\widehat R_{\mathrm{val}}(\hat t)$ (blue),
  the uniform patient-level generalization term $\gamma_{\mathrm{val}}(\delta_{\mathrm{val}})$ (orange; from DKW/Hoeffding),
  and the instability penalty $G_{\mathrm{boot}}$ (red; bootstrap radius passed through the oscillation modulus).
  The black tick marks $R_P(\hat t)$; the number at the bar's right is $U_P(\hat t)$.
  (c) Panel (a) repeated per model (logistic regression, random forest, histogram gradient boosting, AdaBoost, shallow MLP), illustrating model-agnostic application under $P$-frozen regime.}
\end{figure}

\subsection{Notes on figures}

\paragraph{Notes on data-generating processes and algorithmic parameters of figure 1 (illustrative case):}
Internal ($P$) scores are drawn from a two-basin mixture with a "sharp" subgroup (fraction $0.12$) having class-conditional Gaussians $(\mu_0,\mu_1)=(1.9,\,2.1)$ and SDs $(0.15,\,0.13)$, and a "flat" subgroup $(\mu_0,\mu_1)=(2.5,\,4.5)$ with SD $1.0$; $P$ includes heteroskedastic noise (amplitude $2.0$, center $0.90$, width $0.28$). External ($Q$) scores are generated \emph{independently} from their own mixture: "sharp" $(\mu_0,\mu_1)=(1.95,\,2.05)$ with SDs $(0.16,\,0.14)$ and "flat" $(\mu_0,\mu_1)=(2.55,\,4.35)$ with SD $1.05$, plus Q-specific heteroskedastic noise (amplitude $1.8$, center $2.00$, width $0.32$).
Hierarchical sampling for $P$ uses $n_{\text{patients}}=180$ and $800$ cells per patient (so with $n_{\text{internal}}=800$); $Q$ uses $n_{\text{external}}=6000$.
The instability penalty $\Gboot(t)$ uses patient-level bootstrap ($B{=}200$) with moving-average smoothing (window $=5$).
$\lambda$ is calibrated so the $0.58$-quantile of the raw $\Gboot(t)$ matches $1.15\times$ the empirical risk range.
Selection is directional (to the right of $t^{\mathrm{ERM}}$) with movement cap $0.90$ and cost ratio $=1.0$.

%\paragraph{Notes on data-generating processes and algorithmic parameters of figure 2 (illustrative case):}}

\subsection{Additional information on CAMELYON16/17 processing}

\paragraph{Data pre-processing (CAMELYON16/17; slide \& patient levels).}
CAMELYON16 and CAMELYON17 come from different origins, and \emph{both} were processed at two granularities: (i) slide level and (ii) patient level. For each dataset \emph{separately}, we created two disjoint partitions $P$ and $Q$ by a random 50/50 split. Splits were stratified by the task label; at the patient level, all slides from the same patient were kept within the same partition to avoid leakage. Whole-slide images were read at the same fixed magnification, tissue regions were segmented to remove background, and non-overlapping tiles (256--512\,px) were extracted within tissue as preprocessing (see GoogLeNet procedure). Low-quality tiles (low tissue fraction / blur / extreme brightness) were filtered. Each retained tile was scored with a PyTorch GoogLeNet (Inception v1) classifier to obtain $s(x)\in[0,1]$. Slide scores were formed by aggregating tile scores (max); patient scores were then obtained by aggregating a patient's slide scores (quantile). All pre-/post-processing choices (magnification, tiling, QC, aggregators) were held identical in $P$ and $Q$.

\paragraph{Post-threshold reporting (P-frozen).}
Thresholds were selected on $P$ in a selection-honest manner and then \emph{frozen} and evaluated on $Q$. After thresholding, we computed summary performance at both slide and patient levels (e.g., accuracy, risk under $(c_{10},c_{01})$, and related metrics) within each dataset. Owing to space constraints, the main table reports \emph{only mean values} (CAMELYON16 and CAMELYON17 reported separately).

\subsection{Additional information on MIMIC-IV Demo processing}

\paragraph{Data pre-processing (MIMIC-IV-ECG Demo; patient-level split).}
We used only the publicly available MIMIC-IV-ECG \emph{Demo} subset (12-lead, 10\,s, 500\,Hz). To prevent leakage, we randomly split \emph{patients} 50/50 into two disjoint partitions $P$ and $Q$ (all ECGs for a patient remain in the same partition). Each ECG was parsed lead-wise, detrended, and band-pass filtered; amplitudes were $z$-scored per lead (per record). We followed the procedure of \url{https://github.com/nliulab/mimic4ed-benchmark}, with taking the best performing logistic regression. The segment-level classifier produced scores $p_{i,j}\in[0,1]$ for segment $j$ of ECG $i$.

\paragraph{Aggregation and evaluation (P-frozen).}
Per-ECG scores were obtained by aggregating a record's segment scores (e.g., quantile at $q{=}0.9)$: $s_i=g(\{p_{i,j}\}_j)$. Patient-level scores were formed by aggregating across that patient's ECGs (quantile): $S_p=h(\{s_i\}_{i\in \text{patient }p})$. Thresholds were tuned on $P$ in a selection-honest manner and then \emph{frozen} and evaluated on $Q$ (P-frozen). In the main table, we report mean accuracy and mean risk (under $(c_{10},c_{01})$) computed at the selected threshold.

\vfill

\bibliography{thresholder_iastat1}